\documentclass[aps,pra,twocolumn,preprintnumbers,nofootinbib,superscriptaddress,amsmath,amssymb,longbibliography,]{revtex4-2}
\usepackage[T1]{fontenc}
\usepackage{amsmath}
\usepackage{float}
\usepackage{amsfonts}
\usepackage{amssymb}
\usepackage{subfiles}
\usepackage{natbib}
\usepackage[colorlinks,linkcolor=blue,citecolor=blue,breaklinks=true]{hyperref}
\usepackage{graphicx}% Include figure files
\usepackage{dcolumn}% Align table columns on the decimal point
\usepackage{bm}% bold math
\usepackage{color}
\usepackage{booktabs}
\usepackage{amsthm}
\usepackage{dsfont}
\usepackage{tcolorbox}
\usepackage{cases}% http://ctan.org/pkg/cases
\usepackage{physics}
\usepackage{subfiles}

\usepackage{braket}
\usepackage{bbold}
\usepackage{xcolor}
\usepackage[normalem]{ulem} % for crossing out
\usepackage{txfonts,times}
\usepackage{orcidlink}
\usepackage{lipsum}
\usepackage{tabularx}
\usepackage{booktabs}
\usepackage{subcaption}
\usepackage{tikz}         % For overlaying text (the labels)
\usepackage{caption}
\usepackage{comment}
\usepackage{ragged2e}

\begin{document}

\date{\today}

%\title{Characterization of distillable bipartite systems enhanced by collective measurement aided machine learning}
\title{Learning partial-transpose signatures in qubit–ququart states from a few measurements}
\author{Christian Candeago\orcidlink{https://orcid.org/0009-0000-3976-7303}}
\affiliation{Department of Physics, University of Trieste,  34127 Trieste, Italy}
\author{Paolo Da Rold\orcidlink{https://orcid.org/0009-0001-9004-233X}}
\affiliation{Department of Mathematics, Informatics and Geoscience, University of Trieste,  34127 Trieste, Italy}
\author{Michele~Grossi\orcidlink{https://orcid.org/0000-0003-1718-1314}}
\email{michele.grossi@cern.ch}
\affiliation{European Organization for Nuclear Research (CERN), Geneva 1211, Switzerland}
\author{Pawel Horodecki\orcidlink{}}
\affiliation{International Centre for Theory of Quantum Technologies, University of Gdańsk, ul. prof. Marii Janion 4, 80-309 Gdańsk, Poland}
\author{Antonio Mandarino\orcidlink{https://orcid.org/0000-0003-3745-5204}}
\email{antonio.mandarino.work@gmail.com}
\affiliation{International Centre for Theory of Quantum Technologies, University of Gdańsk, ul. prof. Marii Janion 4, 80-309 Gdańsk, Poland}
%\address{Department of Physics  Aldo Pontremoli, University of Milan, Via Celoria 16, 20133 Milan, Italy}

\begin{abstract}
Higher-dimensional quantum systems are attracting interest for improving quantum protocol performance by increasing memory space. Characterizing quantum resources of such systems is fundamental but experimentally costly. We tackle the first non-trivial example: a qubit-ququart system, focusing on partial-transpose spectral classification. Entanglement distillation extracts maximally entangled states from noisy resources, but determining distillability typically requires full state tomography, experimentally prohibitive for high-dimensional systems. We explore a machine learning framework to classify distillable bipartite quantum states using fewer measurements than complete tomography. Our approach employs the PPT criterion, categorizing states by negative eigenvalues in the partial transpose. We use various ML algorithms, including Support Vector Machines, Random Forest, and Artificial Neural Networks, with features from fixed measurements and learnable observables. Results show learnable observables consistently outperform Collective Measurement Witnesses methods. While all models distinguish between non-distillable (PPT) and distillable (NPT) states, differentiating NPT subclasses remains challenging, underscoring the intricate Hilbert space geometry. This work provides an experimentally friendly tool for distillability verification in high-dimensional quantum systems without full state reconstruction.
\end{abstract}

\maketitle

\section{Introduction and Motivation} 
\begin{figure*}
    \centering
    \includegraphics[width=1\textwidth]{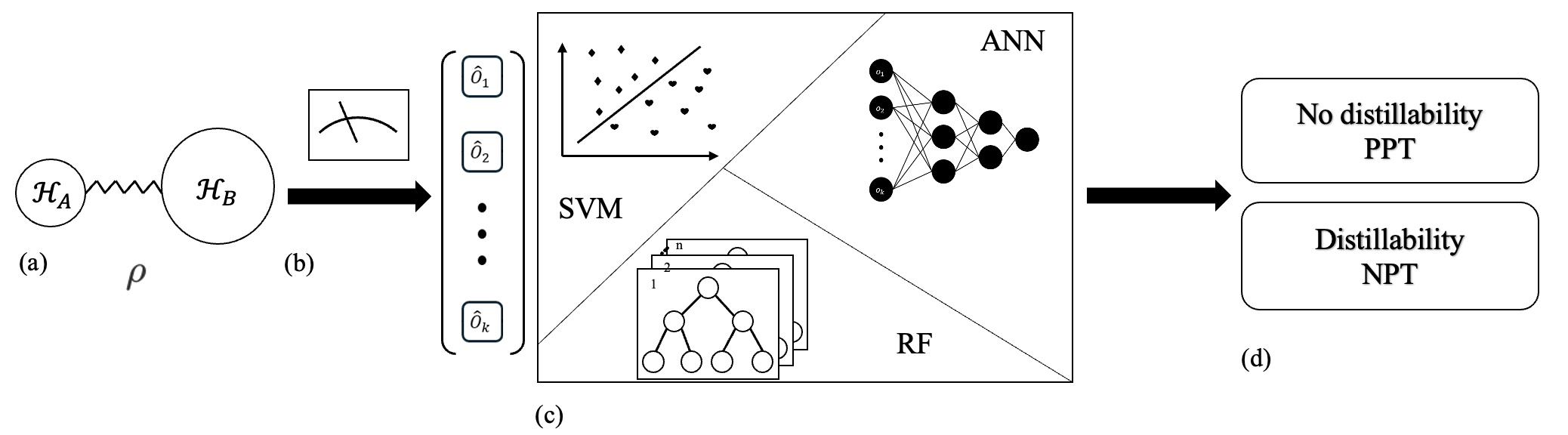}
    \hfill
    \captionsetup{justification=justified, singlelinecheck=false, format=plain, labelfont=bf}
    \caption{\justifying Schematic representation of the machine learning pipeline for quantum state classification. (a) Bipartite quantum state $\rho$ shared between subsystems $\mathcal{H}_A$ and $\mathcal{H}_B$. (b) Measurements performed on the quantum state, yielding observable expectation values as input for learning algorithms. (c) Machine learning classifiers: Support Vector Machine (SVM), Artificial Neural Network (ANN), and Random Forest (RF) trained on the measurement data to distinguish between different entanglement classes. (d) Classification output: states are categorized as either non-distillable (PPT, including separable and bound entangled states) or distillable (NPT, offering resources for entanglement distillation).}
\label{fig:ml_pipeline}
    \label{fig:summary}
\end{figure*}

Quantum entanglement is the fundamental resource that boosts the performance of many quantum technologies over those
of their classical counterparts. However, such an advantage is ascertain and exploitable only in scenarios
dealing with maximally entangled pure states \cite{RMP_Ent,G_hne_2009}. 
In any physical and concrete situation, a prepared copy of an entangled state
undergoes a series of (very often uncontrollable) interactions with the surrounding systems that degrade the amount of such a resource. 
To circumvent such an issue, protocols known as
\emph{entanglement distillation or purification} have been designed that allow one to transform many copies of a noisy entangled state to a target maximally entangled one. 
The first proposals were put forward by Bennett and collaborators with algorithms that work solely on two-qubit states \cite{bennett_purification_1996, Dist_QEC}. There, the authors also showed how entanglement distillation is intertwined with quantum error correcting codes. 
Moreover, in higher-dimensional systems, the probability of successfully distilling a maximally entangled state increases because entanglement can be distributed across a larger number of basis states.\\
The key question now is which states are distillable, and how far distillation can be pushed. The geometry of quantum states living in Hilbert spaces of dimension higher than six makes their characterization problematical with analytical techniques. 
A vast body of literature has explored the conditions and criteria for distillability
with theoretical and computational methods for the case of bipartite systems of dimension $N_A \times N_B$
\footnote{The notation $N_A \times N_B$ is a shorthand to denote that the composite Hilbert space is $\mathcal{H}=\mathcal{H}_A \otimes \mathcal{H}_B \equiv \mathbb{C}^{N_A} \otimes \mathbb{C}^{N_B}$.}, and notable 
example are given in \cite{horodecki_separability_1997, horodecki_mixed-state_1998, dur_distillability_2000, Qubit_qudit_ppt, chen_distillability_2016}.

However, a major challenge is that these criteria typically rely on the full knowledge of the density matrix, which is difficult to obtain experimentally, due to the scaling of the quorum of observables needed to perform a full quantum tomography \cite{Mauro_D_Ariano_2003} as the square of the total dimension (see Section \ref{app:b} for a more thorough discussion). To overcome this limitation and develop a more practical approach for identifying distillable states, we explore the use of machine learning algorithms. 

Recently, machine learning methods have emerged as powerful tools for addressing open, longstanding problems in quantum information science, 
particularly in the detection and quantification of nonlocality and entanglement \cite{Deng2018, balaz2025}. 
For instance, neural networks have been utilized to detect Bell nonlocality in quantum many-body systems, 
providing scalable approaches beyond traditional inequality tests \cite{Zhang2024}. 
Furthermore, machine learning algorithms, including nearest-neighbor classifiers and dimension reduction techniques, 
have been applied to distinguish bound entangled states, those undetectable 
by the PPT criterion, from separable states in high-dimensional bipartite qutrit systems, 
revealing significant volumes of bound entanglement and novel substructures within magically symmetric state spaces \cite{Hiesmayr2021}. 
These advancements highlight the potential for ML to tackle hard challenges necessary for  
advancing quantum technologies reliant on reliable nonlocality and entanglement verification.

We focus here on learning schemes capable of classifying the number of negative eigenvalues of the partially transposed density matrix describing the system. 
The Positive Partial Transpose (PPT) criterion provides a necessary condition for separability: if a bipartite state $\rho$ is separable, its partial transpose $\rho^\Gamma$ must be positive semidefinite. Thus, the presence of negative eigenvalues in $\rho^\Gamma$ implies that $\rho$ is entangled.
This criterion is also a necessary condition for entanglement distillability \cite{horodecki_inseparable_1997}. 
For dimensions beyond $2 \cross 3,$ NPT does not provide a complete characterization of distillability \cite{kraus_2000}.
Throughout, we focus on classifying NPT structure (i.e., the number of negative eigenvalues) as an experimentally accessible surrogate for distillation usefulness.
A state is said to be distillable if maximally entangled pairs can be extracted from multiple copies of it using LOCC. All NPT states in $2 \times 2$ and $2 \times 3$ dimensions are known to be distillable. In higher dimensions, such as $2 \times 4$, the existence of one or more negative eigenvalues in $\rho^\Gamma$ suggests distillability, though the full characterization remains open. Conversely, PPT entangled states are bound entangled and cannot be distilled.

%some statement about potential applications
% NEED TO ADD SOME REFERENCES
\subsection{Operational motivation and applications}
One of the most immediate and compelling applications lies in the development of quantum repeaters for long-distance quantum communication.
The distribution of entanglement across large scales is severely hindered by channel loss and environmental noise, leading to rapid degradation of the entangled resource.
Quantum repeaters circumvent this limitation by combining entanglement swapping and entanglement distillation at intermediate nodes, thereby extending the communication range, see e.g.\ the review paper \cite{RevModPhys_rep}.
In a realistic repeater architecture, a node must repeatedly decide \emph{on the fly} how to process each newly generated elementary link: (i) perform a small number of local measurements to obtain a compact feature record, (ii) use this record to infer whether the shared state is compatible with being a useful input for a purification/distillation routine, and then (iii) either (a) discard the link and request a new attempt if it is predicted to be non-distillable (PPT-like), or (b) route it to the next stage of the protocol, for instance, run a distillation step and, if successful, proceed to entanglement swapping with neighboring links.
In this setting, a lightweight classifier that flags PPT versus NPT behavior (and, more finely, provides an NPT-subclass label) can serve as a decision module that reduces the need for full state tomography at intermediate stations, lowering experimental overhead while enabling adaptive resource management in repeater-based networks \cite{Torta2025}.

A second domain where high-dimensional entanglement has attracted considerable attention is quantum key distribution (QKD).
Encoding information in larger Hilbert spaces not only increases the key rate but also enhances resilience to noise and improves robustness against certain classes of eavesdropping strategies \cite{Erhard_2020}.
However, these benefits revolve around the ability to generate reliably and \emph{validate} high-dimensional entanglement under realistic laboratory conditions.
Here, a concrete decision flow is again natural: after distributing candidate entangled pairs, the parties perform a limited set of measurements to compute features, apply a trained classifier to rapidly screen blocks that are unlikely to be useful (e.g.\ PPT-like blocks), and prioritize parameter-estimation \cite{Lupo2018}, error-reconciliation \cite{Buttler_2003}, or entanglement-based post-processing only on the blocks predicted to retain distillation-relevant correlations \cite{Huang2014}.
Such a screening step can reduce wasted acquisition time and help stabilize operation in regimes where full characterization would be too costly, thereby supporting next-generation QKD protocols that exploit high-dimensional resources \cite{Mirhosseini_2015}.

The relevance of distillation also extends into the domain of quantum error correction (QEC), which underpins the feasibility of large-scale, fault-tolerant quantum computation \cite{Rengaswamy2024entanglement}.
Hybrid qubit--qudit systems, such as those studied here (and their multipartite generalizations), are a natural setting for fault-tolerant quantum information processing protocols, since they allow one to treat some physical subsystems as effective qubits while exploiting higher-dimensional levels to implement nonbinary stabilizer quantum error-correcting codes and syndrome-extraction architectures tailored to qudits.
From an operational viewpoint, the same ``measure--decide--act'' logic applies: limited measurement data can be used to quickly decide whether a noisy distributed resource should be fed into a purification preprocessing step, stored for later use, or discarded in favor of a fresh preparation attempt.
In this perspective, a measurement-efficient classifier that detects distillation-relevant signatures can be viewed as a complement to QEC-based architectures, enabling adaptive strategies for resource preliminary assessment and potentially improving the efficiency of high-dimensional protocols \cite{QEC_qudit}.

Finally, beyond direct technological applications, there is considerable foundational interest in understanding the geometry of entangled states in higher dimensions.
Low-dimensional entanglement has been extensively characterized \cite{Lu_2018}, but as the system size grows the structure of the state space becomes increasingly intricate, giving rise to new classes of bound entangled states and complex separability boundaries.
By applying machine learning to the classification of distillation-relevant structure, one does more than address a practical challenge: one also opens the door to uncovering hidden geometric and structural patterns in the space of quantum states.
Insights gained from such an exploration may enrich our theoretical understanding of entanglement and inform the design of future protocols that explicitly leverage high-dimensional quantum correlations \cite{Urena2024}.\\

This work focuses on bipartite states of dimension \(2\times 4\) \footnote{In a \(2\times 4\) bipartite quantum system, bound entangled states with positive partial transpose can hinder distillation into maximally entangled forms like the GHZ or W states of 3 qubits, as they resist conversion via local operations and classical communication despite sharing the same 8-dimensional Hilbert space.}.
We introduce a machine-learning-based framework that classifies PPT versus NPT behavior and it opens the way to tackle the characterization of NPT subclasses in \(2\times N\) systems without requiring complete state reconstruction. Our contributions are threefold: (i) we introduce a measurement-efficient supervised classifier for qubit--ququart (\(2\times 4\)) states that distinguishes PPT from NPT instances and, more finely, predicts the NPT class defined by the number $\xi$ of negative eigenvalues of the partial transpose; We use three learning approaches: Artificial neural network (ANN), Support Vector Machine (SVM)\cite{Vapnik1995} and Random Forest (RF) \cite{Breiman2001};  (ii) with ANN we benchmark standard fixed collective-measurement-witness (CMW) features against a learnable-observable layer in which the measured Hermitian operators are optimized jointly with the network parameters; and (iii) we provide empirical evidence that separating the NPT\(_1\) and NPT\(_2\)\footnote{NPT$_\xi$ defines the class by the number of negative eigenvalues of the partial transpose, see Sec \ref{sec:results}} subclasses is intrinsically difficult, consistent with a strong overlap of these classes in the feature space and the resulting lack of a simple decision boundary.
We graphically summarize our testing scheme in Fig.\ref{fig:summary}. 
Eventually, due to the emergence of qudits' practical relevance in experimental platforms \cite{genqdit, genqdit1, genqdit2,genqudit3}, such as angular momentum of photons, trapped ions, and superconducting artificial atoms, our approach paves the way to an ML-enhanced method to distill other nonclassical resources, such as contextuality and magic. 

\begin{figure}[h!]
    \centering

    \begin{subfigure}{0.5\textwidth}
        \captionsetup{labelformat=empty}
        \begin{tikzpicture}[remember picture]
            \node[inner sep=0] (f1ave) {\includegraphics[width=\linewidth]{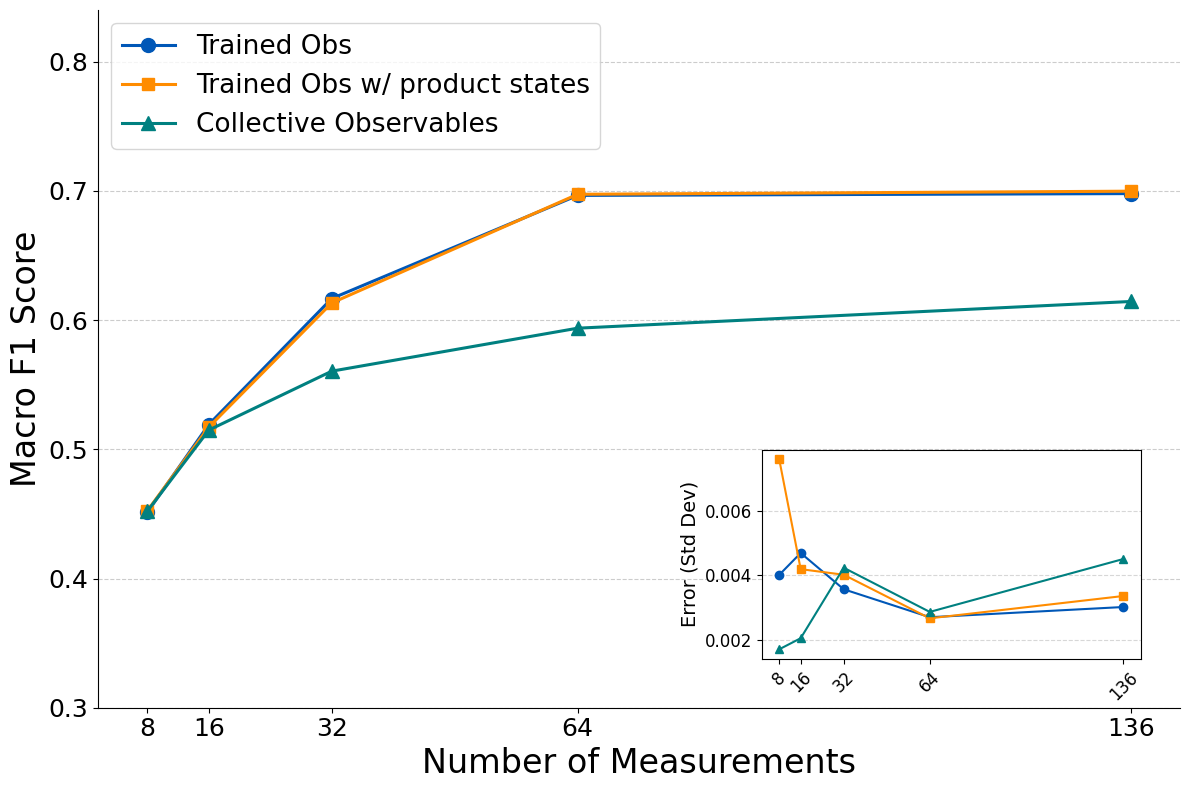}};
            \node[anchor=south west, xshift=5pt, yshift=5pt, fill=white, fill opacity=0.7, text opacity=1, inner sep=1pt] at (f1ave.south west) {\subref*{fig:f1_ave}};
        \end{tikzpicture}
        \caption{} % Needed to increment subfigure counter
        \label{fig:f1_ave}
    \end{subfigure}
    \hfill
    \begin{subfigure}{0.49\textwidth}
        \captionsetup{labelformat=empty}
        \begin{tikzpicture}[remember picture]
            \node[inner sep=0] (f1class) {\includegraphics[width=\linewidth]{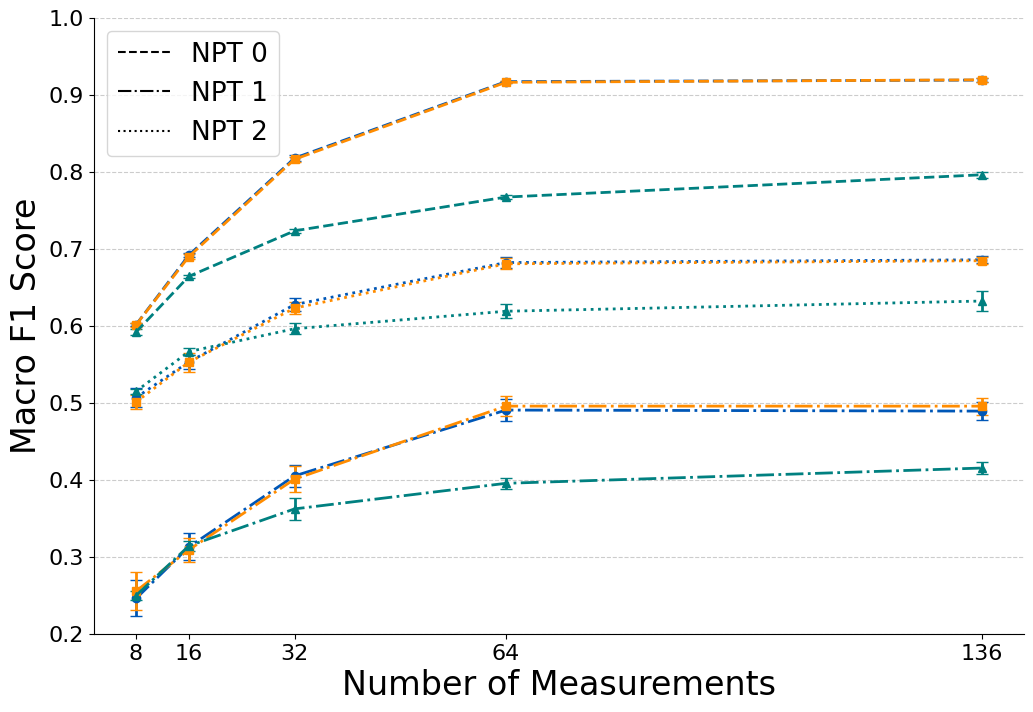}};
            \node[anchor=south west, xshift=5pt, yshift=5pt, fill=white, fill opacity=0.7, text opacity=1, inner sep=1pt] at (f1class.south west) {\subref*{fig:f1_class}};
        \end{tikzpicture}
        \caption{}
        \label{fig:f1_class}
    \end{subfigure}

    \hfill
    \captionsetup{justification=justified, singlelinecheck=false, format=plain, labelfont=bf}
    \caption{\justifying Summary of the results obtained with the different models as a function of $k$. All plots are generated using $10^6$ quantum states, split into training and test sets, and each data point represents the average over 10 independent training runs.
    (a) Mean Macro-F1 score, defined in Eq.~\eqref{eq:Macro-F1}, as a function of $k$. Models employing learnable observables achieve comparable performance and consistently outperform the model based on CWM. Insets show the corresponding standard deviations.
    (b) Class-wise F1 scores, with colors and markers matching those in panel (a). The model performance is lowest for the NPT$_1$ and NPT$_2$ classes, whereas NPT$_0$ states are correctly classified in the majority of cases.}
    \label{fig:f1_plot}
\end{figure}
\section{Results}
\label{sec:results}
It is known that under the PPT criterion, we have at most 3 negative eigenvalues for a qubit-ququart state \cite{rana_negative_2013}. Since a higher number of negative eigenvalues typically indicates a greater amount of entanglement available for distillation, knowing how many negative eigenvalues a state has is crucial for assessing its distillability.
To apply the PPT criterion, the quantum state must be fully characterized, i.e., its density matrix ${\rho}$ must be known. This requires performing full quantum state tomography, which is particularly resource-intensive. For large systems, the number of required measurement runs scales as $O(N_A N_B)$, rapidly rendering the procedure experimentally infeasible.
To address this challenge, we built an analysis based on machine learning schemes, inspired by the approach introduced in \cite{travnicek_sensitivity_2024}, to investigate the distillability of bipartite quantum states of dimension $2 \times 4$ using significantly fewer measurements than those required for full state tomography.
This information can be used as a practical screening tool for deciding whether a distributed state is compatible with being a useful input to purification/distillation routines, without claiming a universal ``distillability oracle'' for the full \(2\times 4\) state space.

A key takeaway is that our scheme should be interpreted as a resource-efficient diagnostic of partial-transpose structure, not as a definitive test of distillability in \(2\times 4\). The ability to reliably separate PPT states (which are undistillable) from NPT states is already operationally valuable in scenarios that require rapid triage of distributed entangled resources, and the predicted NPT subclass provides additional coarse-grained information about the ``strength'' of the NPT signature. At the same time, because NPT is not sufficient for distillability in higher dimensions, connecting our learned labels to the performance of specific distillation protocols (e.g.\ yields or thresholds under concrete LOCC procedures) remains an important direction for future work. In this sense, the present approach provides a lightweight, experimentally motivated screening layer that can be composed with protocol-specific analyses, rather than replacing them.

We categorize the quantum states according to the number of negative eigenvalues of their partial transpose. Since only non-positive partial transpose (NPT) states are distillable, we restrict our analysis to states exhibiting one or more negative eigenvalues. For the sake of convenience, in the following, we will call a PPT state NPT$_0$ (meaning zero negative eigenvalues under partial transposition), a state with one negative eigenvalue of the partial transpose NPT$_1$, and similarly for the other classes of states. The states used for the classification task are sampled randomly (see the Methods section \ref{app:data} for details on the generation of the states), with NPT$_1$ and NPT$_2$ being by far the most typical states obtained. Thanks to that, we limit our classification to NPT$_\xi$ states, with $\xi \in \{0, 1, 2\}$, as NPT$_3$ states are not typical. 
Shortly, we decided to omit NPT$_3$ states because they do not provide much more information at the cost of being much harder to sample. Our analysis is based on random quantum states sampled from different probability measures. It is well known that the space of density matrices does not admit a unique, physically distinguished probability measure. The sampling according to the Hilbert-Schmidt and Bures measures resulted in approximately one NPT$_3$ state over $ 10^6$. Therefore, we assumed them not to be typical random states \cite{zyczkowski2001induced, zyczkowski2011generating}. A detailed analysis is reported in Table~\ref{tab:random_states} in Sec.~\ref{sec:2x4_pt_neg}. 

We employ three different neural network frameworks for classifying the states, alongside SVM and RF. We start with an ANN whose input features are derived from a fixed set of observables known as Collective Measurement Witnesses (CMW), which constitute a well-established set of observables for entanglement classification. In contrast, the second and third ANN models do not rely on a predefined feature set; instead, they learn the optimal observables during the training process. This adaptive approach allows for the discovery of a more expressive set of operators. Although the second model operates on a single state $\rho$, the third utilizes several copies of the same state, i.e., the expectation values of the observables are computed on the tensor product of states $\rho^{(l)}$, see Eq.~\eqref{eq:dm_cwm}. These higher-order moments of the density matrix provide deeper insights into the system's non-classicality \cite{PRL_2003}. For all three models, the number of observables $k$ remains a critical hyperparameter. Detailed explanations of these techniques are provided in Sections \ref{app:cmw} and \ref{app:obs}.\\
Concerning SVM and Random Forests RF, we followed the same learning protocol adopted for the first ANN model. For each fixed number $k$ of input observables, model hyperparameters were optimized through a grid-search procedure over a predefined parameter space, ensuring a consistent comparison across different measurement budgets. For both approaches, we used 4,500 training samples evenly distributed across the classes and a test set of 1,500 samples.\\
To study the performance of the considered models, we employed the Macro-F1 score as it offers a simple metric for cases where we are interested in both the precision and the recall. 
For a classification problem with $C$ classes, the Macro-F1 score is given by the arithmetic mean of the F1-scores computed for each class:
\begin{equation}
\label{eq:Macro-F1}
\mathrm{Macro\text{-}F1} = \frac{1}{C} \sum_{c=1}^{C} \mathrm{F1}_c,
\end{equation}
where the F1-score of class $c$ is defined as the harmonic mean of Precision and Recall:
\begin{equation}
\mathrm{F1}_c = \frac{2\,\mathrm{Precision}_c \cdot \mathrm{Recall}_c}
{\mathrm{Precision}_c + \mathrm{Recall}_c},
\end{equation}
and we recall that:
\begin{equation}
\mathrm{Precision} = \frac{\mathrm{TP}}{\mathrm{TP} + \mathrm{FP}},
\quad
\mathrm{Recall} = \frac{\mathrm{TP}}{\mathrm{TP} + \mathrm{FN}},
\end{equation}
where TP, FP, and FN denote the number of true positives, false positives, and false negatives, respectively.
This metric assigns equal importance to all classes, independently of their relative frequencies, and is therefore particularly suitable for imbalanced classification problems.\\

As expected, the number of observables used for state classification, whether CMW or learned observables, has a direct impact on the model performance. In particular, for ANN models, increasing the number of observables $k$ leads to improved accuracy, up to $k = 64$ after which it saturates. As depicted in Fig. \ref{fig:f1_ave}, gains beyond this point are minimal, meaning that more observables are useless, as known from state tomography. 

\begin{figure*}
    \centering
    \includegraphics[width=1\textwidth]{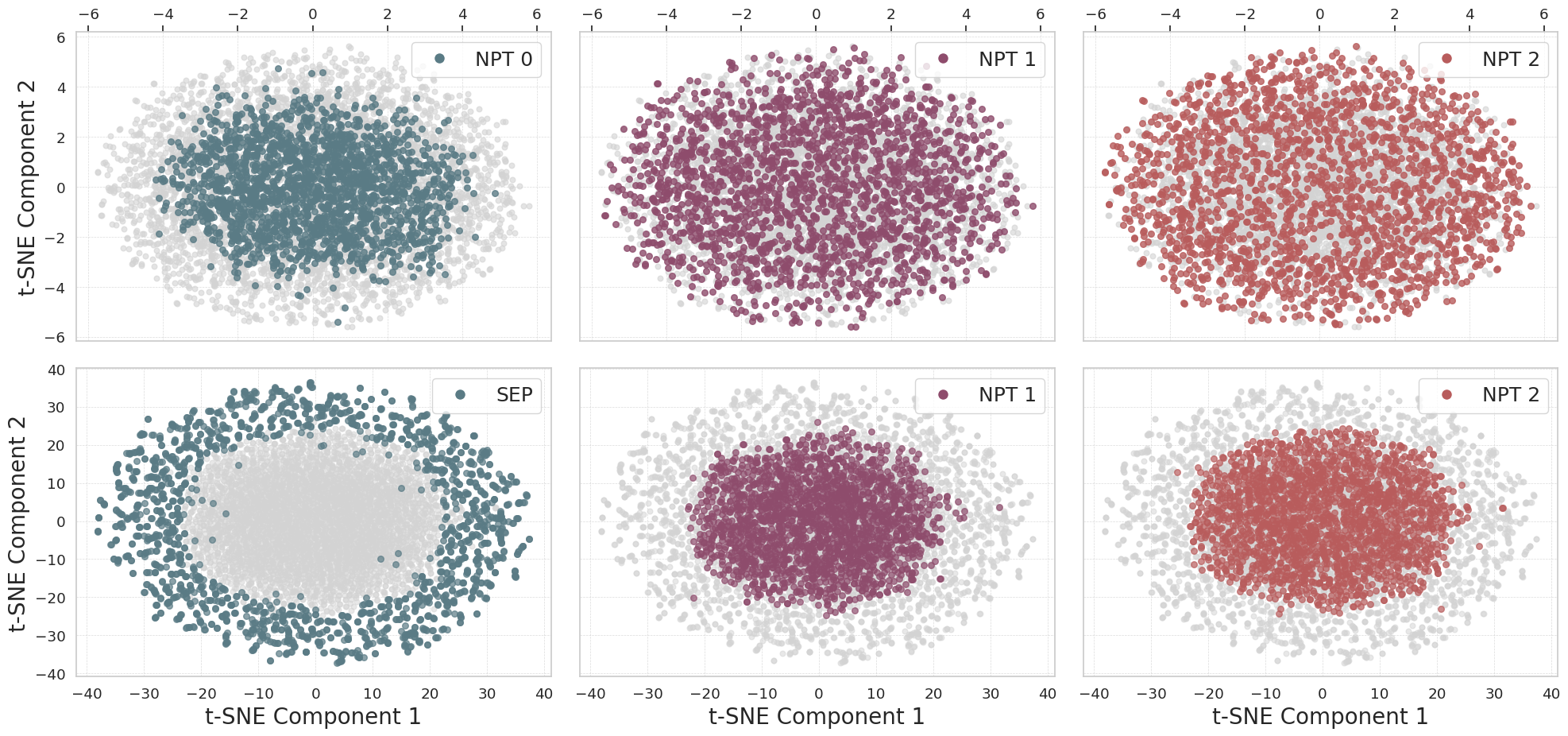}
    \hfill
    \captionsetup{justification=justified, singlelinecheck=false, format=plain, labelfont=bf}
    \caption{\justifying Plots representing the t-SNE of the expectation values of 64 learned observables for different classes, each made up of $10^3$ different states. 
    Each subplot highlights a specific class to improve readability, though all points are part of the same t-SNE embedding. In the top row the embedding includes NPT$_0$, NPT$_1$ and NPT$_2$ states. These classes appear largely intertwined, although the NPT$_0$ shows some clustering. 
    On the bottom row we changed the NPT$_0$ states with separable product states. In this case, while NPT$_1$ and NPT$_2$ remain heavily mixed, the product states forms a well separated cluster.}
    \label{fig:tsne}
\end{figure*}

Fig. \ref{fig:f1_ave} strikingly reveals a key strength of our approach: for any fixed number $k$ of observables, models with learnable observables consistently outperform those relying on fixed CMW measurements—demonstrating the power of data-driven observable optimization. While overall classification accuracy remains moderate across models (highlighting the intrinsic challenge of NPT subclass resolution), Fig. \ref{fig:f1_class} provides critical class-wise insight into the Macro F1 scores, exposing where our methods excel and pinpointing geometric bottlenecks for future refinement.
We observe a high classification accuracy for NPT$_0$ states with a value of Macro F1 score around $0.75-0.90$, moderate performance for NPT$_2$ states with values of Macro F1 in the range of $0.60-0.70$, and poor performance for NPT$_1$ states with values of Macro F1 lesser than $0.50$.
We remark that these observations are independent of the model used for the classification.\\
In tab.~\ref{tab:macro-f1-all-models} are reported all the numerical results for the Macro-F1 score for the three ANN models, SVM, and RF. It's straightforward to see that SVM and RF are the models that give the worst results.\\
To rule out limitations in our models we reformulated the task as a binary classification problem and tried to train the models with two different datasets. In the first dataset the NPT$_0$ states were replaced in favor of product states of the form $\rho = \rho_A \otimes \rho_B$. The model achieved high accuracy in fewer epochs, indicating that this classification is trivial. The second dataset was made with only NPT$_1$ and NPT$_2$ states. In this case the model has a hard time discerning the two classes, with accuracy oscillating around 50\% , indicating that it struggles to learn a meaningful decision boundary.
To substantiate our findings, we employed the t-Distribuited Stochastic Neighbour Embedding (t-SNE) \cite{tsne}, which can map the high-dimensional vector of expectation values to a more understandable 2-dimensional plot. The results are presented in Figure \ref{fig:tsne}, and further validate our findings: NPT$_0$ states are more distinguishable, and product states even more so, whereas other states exhibit significant overlap, rendering them inherently difficult to classify.

\subsection{Intrinsic difficulty in separating NPT$_1$ and NPT$_2$ classes} 

A fundamental aspect emerging from our analysis is that the poor separability between the NPT$_1$ and NPT$_2$ classes does not appear to be an artefact of the chosen machine-learning models, but rather an intrinsic feature of the geometry of the $2 \times 4$ state space.
All the approaches, that we employed, show a clear difficulties to learn the geometry of the subspace of entangled states, since the classification of NPT$_1$ and NPT$_2$ states lead to a random guesser behaviour (accuracy around 50\%) as reported in Table~\ref{tab:macro-f1-all-models}. This intrinsic difficulty is also confirmed by the analysis of the Hilbert space using the t-SNE projection in Fig.~\ref{fig:tsne}. Here we observe that the regions in the Hilbert space corresponding to NPT$_1$ and NPT$_2$ states are heavily intertwined with respect to the considered sets of observables. This makes extremely complicated the search of a boundary between the two subspaces within the fixed measurement budget. A more in-depth analysis about this difficulty is in Sec.~\ref{sec:svd}.
Our results suggest that, for realistic measurement schemes with a limited number of global observables, the spectral structure of the partial transpose alone does not induce NPT$_1$/NPT$_2$ labels that are easily learnable, highlighting a fundamental geometric limitation on the resolution of few-data classification protocols.

\begin{table}[t]
\centering
\caption{ \justifying Macro-F1 results of the main learning models used for the problem, namely artificial neural network (ANN) in the three different framework, Support Vector Machine (SVM), and Random Forest (RF).}
\label{tab:macro-f1-all-models}
\begin{tabular}{lrrrrr}
\hline
Macro-F1 Score (N. obs $k$) &  8 & 16 & 32 & 64 & 136 \\
\hline
ANN - Trained obs & 0.45 &  0.52  & 0.62 & 0.70 & 0.70 \\
ANN - Trained obs w/ product states & 0.45 &  0.52  & 0.61 & 0.70 & 0.70 \\
ANN - CWM & 0.42 &  0.51 &  0.56 &  0.59 & 0.61 \\
SVM & 0.43 & 0.47 & 0.52 & 0.58 & 0.59 \\
RF & 0.43 & 0.46 & 0.47 & 0.51 & 0.51 \\

\hline
\end{tabular}
\label{tab:macro-f1-all-models}
\end{table}

\section{Discussion}
The empirical success of our measurement-efficient classifiers, despite inherent challenges in resolving NPT$_1$ from NPT$_2$ subclasses, 
underscores the need for alternatives to full quantum state tomography, which scales poorly with system dimension. 
Here, we analyze this tomographic overhead and contrast it with the measurement budgets of our ML approach, 
highlighting its feasible implementation, motivated by task-tailored learning schemes that bypass complete state reconstruction, 
while it reliably identifies a state offering resources for purification protocols.

\subsection{Scaling of a tomographic reconstruction of a quantum state}
\label{app:b}
We recall that standard quantum state tomography requires an informationally complete \emph{quorum}, i.e., a spanning set of effects whose expectation values suffice to reconstruct \(\rho\), which leads to an experimental effort scaling as \(\Theta(d^2)\) in the Hilbert-space dimension ($d\equiv N_A \times N_B$). 
A convenient way to make this scaling explicit is to expand \(\rho\) in an orthonormal operator basis \(\{B_j\}_{j=1}^{d^2-1}\) of traceless Hermitian operators, \(\mathrm{Tr}(B_j)=0\) and \(\mathrm{Tr}(B_jB_k)=\delta_{jk}\), so that:
\begin{equation}
\rho = \frac{\mathbb{I}}{d} + \sum_{j=1}^{d^2-1}\theta_j B_j,
\qquad \theta\in\mathbb{R}^{d^2-1}.
\end{equation}
Consider a collection of two-outcome measurements \(\{P_k,\mathbb{I}-P_k\}\) with:
\begin{equation}
P_k=\frac{\mathbb{I}}{d}+\sum_{j=1}^{d^2-1}p_{kj}B_j .
\end{equation}
Then the Born rule becomes a linear model in the parameters:
\begin{equation}
f_k := \Pr(1|\rho,P_k)=\mathrm{Tr}(\rho P_k)=\frac{1}{d}+p_k^\top \theta,
\qquad
f=\frac{1}{d}+X\theta,
\end{equation}
where \(X_{kj}=p_{kj}\).  A quorum corresponds to choosing \(\{P_k\}\) such that the induced linear map has full rank (informational completeness), which in turn requires on the order of \(d^2\) independent constraints \cite{Mauro_D_Ariano_2003,Granade_2016}.
For $d-$dimensional systems, the number of independent real parameters and the size of a generic quorum scale as $d^2$
rendering full tomography (and thus direct evaluation of tomography-based distillability tests) exponentially costly in the dimension of the system \cite{Granade_2016}. 
Several strategies reduce the effective measurement budget by replacing generic tomography with \emph{structured} tomography or physics-informed one \cite{Fid_recon_tomo}.
A prominent example is compressed-sensing tomography presented in \cite{Gross_2010}, which leverages approximate low rank: if \(\rho\) has rank \(r\ll d\), then one can reconstruct \(\rho\) from only
\begin{equation}
m = O\!\left(r d \log^2 d\right)
\end{equation}
randomly chosen Pauli expectation values, rather than \(O(d^2)\) settings.
Operationally, one measures expectations \(\mathrm{Tr}(\rho\,w(\Sigma_i))\) of randomly sampled \(n\)-qubit Pauli strings \(w(\Sigma_i)\), and reconstructs \(\rho\) via a convex program such as trace-norm minimization,
\begin{eqnarray}
\min_{\sigma}\ \|\sigma\|_{\mathrm{tr}}
\, \, \text{s.t.}\,
\mathrm{Tr}(\sigma)=1,\,
\mathrm{Tr}\!\left(w(\Sigma_i)\sigma\right)=\mathrm{Tr}\!\left(w(\Sigma_i)\rho\right),
\end{eqnarray}
with $\ (i=1,\dots,m)$ and robustness guaranties under noise and approximate low rank.
However, low rank states are not typical in the Hilbert space, and this justifies the efforts to develop ML-aided strategies for the characterization of quantum resources.  

This is due to the observation that to target full-rank tomography, the \emph{choice} of the quorum strongly affects statistical efficiency and numerical stability.
In the linearized model \(Y=X\theta\) (with empirical frequencies \( \hat f \) and \(Y=\hat f-1/d\)), reconstruction by least squares yields:
\begin{equation}
\hat\theta_{\mathrm{LS}} = (X^\top X)^{-1}X^\top Y,
\end{equation}
and the sensitivity to noise is controlled by the conditioning of \(X\).  A standard diagnostic is the condition number:
\begin{equation}
\kappa(X)=\frac{\sigma_{\max}(X)}{\sigma_{\min}(X)},
\end{equation}
where \(\sigma_{\max}\) and \(\sigma_{\min}\) are the largest and smallest singular values, small \(\kappa(X)\) corresponds to well-conditioned (noise-robust) inversion, whereas ill-conditioned designs amplify statistical fluctuations \cite{Granade_2016}.
On this line, we analyze how singular value decomposition can be integrated into the feature selection in Sec.~\ref{sec:svd}. 

This motivates \emph{adaptive} quorum design, where one selects new measurement settings based on previous data so as to maximize information gain or otherwise minimize expected posterior uncertainty.
In Bayesian adaptive tomography, one formalizes this by assigning a utility \(U(X)\) to a prospective measurement design \(X\), e.g.
\begin{equation}
U(X)=\mathbb{E}_{\hat f,\theta|X}\!\left[L(\theta)\right],
\end{equation}
for a chosen loss function \(L\), and then choosing the next setting to optimize \(U\), updating the posterior iteratively as data are collected.
Practical schemes combine lightweight online heuristics for selecting the next measurement with heavier offline estimators (Bayesian mean estimation, maximum likelihood, or constrained least squares), enabling improved accuracy without the need to predefine a fixed tomographically complete quorum in advance. \cite{Granade_2016,Mauro_D_Ariano_2003}

To finish, we like to outline some implications for distillability detection and the motivation that pushed the work presented. 
The considerations presented in this subsection motivate replacing full-state reconstruction (requiring \(\Theta(d^2)\) informational degrees of freedom) by protocols that either (i) exploit structure to reduce the effective measurement scaling to sub-quadratic in \(d\) (e.g., \(O( d\log d)\) ), or (ii) bypass tomography entirely by learning task-specific decision rules from incomplete measurement records. Accordingly, we examined machine-learning classifiers trained on low-dimensional feature vectors extracted from observables (\(m\ll d^2\)), to identify distillable states at a fraction of the measurement cost required by full tomography. 

\subsection{Enhanced learning by collective measurement aided machine learning}
For bipartite systems beyond \(2\times 3\), the negative partial transpose (NPT) property is best viewed as a \emph{distillation-relevant} signature rather than a complete characterization of entanglement distillability. In particular, while PPT entangled states are undistillable (bound entangled), NPT is in general only a necessary condition for distillability in higher dimensions and does not, by itself, define an operational distillation protocol. Throughout this work, we therefore adopt a deliberately conservative standpoint: our goal is to learn, from limited measurement data, whether a given state exhibits PPT or NPT behavior and, more finely, to predict the NPT subclass defined by the number \(k\) of negative eigenvalues of the partial transpose. 
Rather than providing a universal ``distillability oracle'' for the full \(2\times 4\) state space, this information serves as a practical screening module that actively flags which distributed states are promising candidates for purification/distillation routines and which ones can be safely discarded.
In this study we addressed the problem of distinguishing $2\times4$-dimensional quantum states based on the number of negative eigenvalues in their partial transpose, as determined by the Positive Partial Transpose (PPT) criterion. For the sake of avoiding full quantum state tomography, we employed the use of machine learning to classify the states using the expectation values of some observables. We explored three such models: one using a fixed set of CMWs, and two others where the observables are integrated into the network as learnable parameters.\\
Our results demonstrate how machine learning can help distinguish non-distillable states (NPT$_0$) from distillable states (NPT$_1$ and NPT$_2$). We also found that the model utilizing learnable observables consistently outperforms the one based on CMWs measures and that the accuracy gains beyond 64 observables are minimal. Crucially, this measurement saturation aligns with data-informed strategies in adaptive tomography, where new observables are dynamically selected based on prior data to maximize efficiency as reported in Sec. \ref{app:b}.
While the proposed protocol successfully identifies distillable states, the distinction between NPT$_1$ and NPT$_2$ states remains challenging for all considered models. 
These outcomes, together with t-SNE plots and binary classification tasks, indicate that the observed difficulty is unlikely to stem from a limitation of our method, and instead appears to be a consequence of the particularly complex geometric structure of quantum states in the relevant dimension. \\
As mentioned before, this work illustrates how learnable observables may prove better for these types of tasks and could be an interesting approach to different problems. While the experimental realization of such observables still requires careful assessment, their flexibility enables the discovery of optimal solutions beyond human intuition, leading to more efficient data-driven approaches in quantum information. However, a possible experimental implementation may be achieved via standard dilation schemes, which require an increase in the total Hilbert space dimension through the addition of auxiliary subsystems that are ultimately traced out.\\
%From this work, several open problems emerge following the difficulties encountered in separating NPT$_1$ and NPT$_2$ states. We observe that the geometry of the Hilbert space of bipartite quantum states in dimension $2\times4$ is nontrivial, particularly within the subspace of entangled states. This raises the question of whether a classical approach can effectively characterize the structure of this subspace, or the problem is intrinsically hard, such that only a feasible quantum learning approach could provide meaningful insight. This could be a future direction of research and application for quantum algorithms. Furthermore, we can ask whether it is possible to establish a mapping between the space of bipartite states in dimension $2\times4$ and the space of tripartite states in dimension $2\times2\times2$. 
Our findings naturally lead to several transformative open questions, rooted in the fundamental difficulty of discriminating between NPT$_1$ and NPT$_2$ states, which reflects the highly nontrivial geometric structure of $2 \times 4$ bipartite entanglement. 
This nontrivial Hilbert space structure, particularly within the entangled subspace, poses a fundamental question: can classical machine learning ever fully map its intricate topology, or does it demand intrinsically quantum-enhanced learning paradigms to unlock meaningful characterizations? Such quantum machine learning (QML) architectures could herald breakthroughs in algorithms for distillation of quantum resources, setting a new paradigm for high-dimensional resource verification.

Looking ahead, our findings open several ambitious avenues for ML-enhanced protocols. An application is to establish engineered strategies that can exploit the transition between $2 \times 2N$ bipartite systems and $2 \times \underbrace{2 \times\dots \times 2}_{N-times}$ multipartite ones, such as potentially reinterpreting the degrees of freedom of the ququart as the atomic 4-level $\Lambda\!\!-\!\!V$ systems of two interacting qubits in alkali compounds. On a more foundational side, learning strategies for the brachistochrone dynamics can be developed that interpolate between known NPT$_1$ and NPT$_2$ states, enabling predictive control over partial-transpose evolution and, lastly, developing tailored metrics, such as quantum geometric tensors or Fubini-Study distances, to quantify Hilbert space geometry and separability boundaries. Finally, a most natural generalization is to put forward a scalable generalization to $2\times N$  systems, where hybrid QML protocols could seriously push a speed-up for distillation protocols for quantum networks, repeaters, and fault-tolerant computing. 

\section{Methods}
\subsection{Data generation}
\label{app:data}
\begin{figure}[h!]
    \centering

    \begin{subfigure}{0.5\textwidth}
        \captionsetup{labelformat=empty}
        \begin{tikzpicture}[remember picture]
            \node[inner sep=0] (f1ave) {\includegraphics[width=0.95\linewidth]{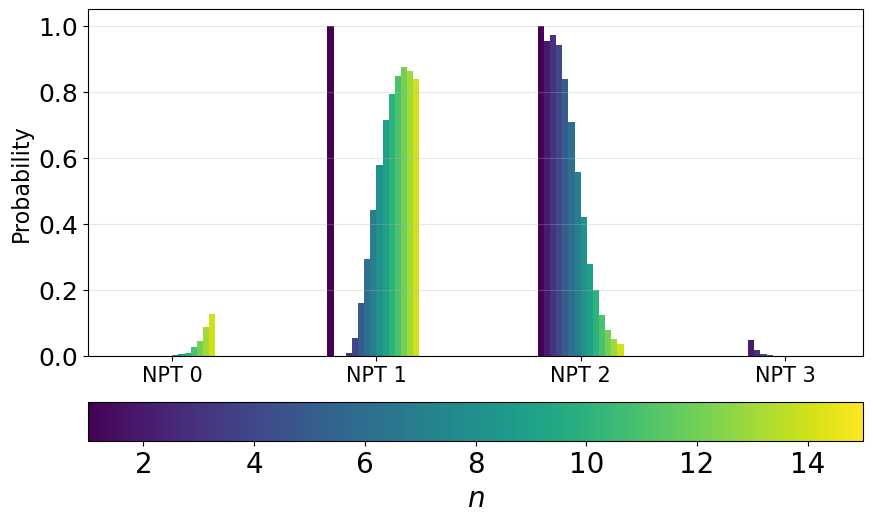}};
            \node[anchor=south west, xshift=5pt, yshift=5pt, fill=white, fill opacity=0.7, text opacity=1, inner sep=1pt] at (f1ave.south west) {\subref*{fig:hist_mix}};
        \end{tikzpicture}
        \caption{} % Needed to increment subfigure counter
        \label{fig:hist_mix}
    \end{subfigure}
    \hfill
    \begin{subfigure}{0.485\textwidth}
        \captionsetup{labelformat=empty}
        \begin{tikzpicture}[remember picture]
            \node[inner sep=0] (f1class) {\includegraphics[width=\linewidth]{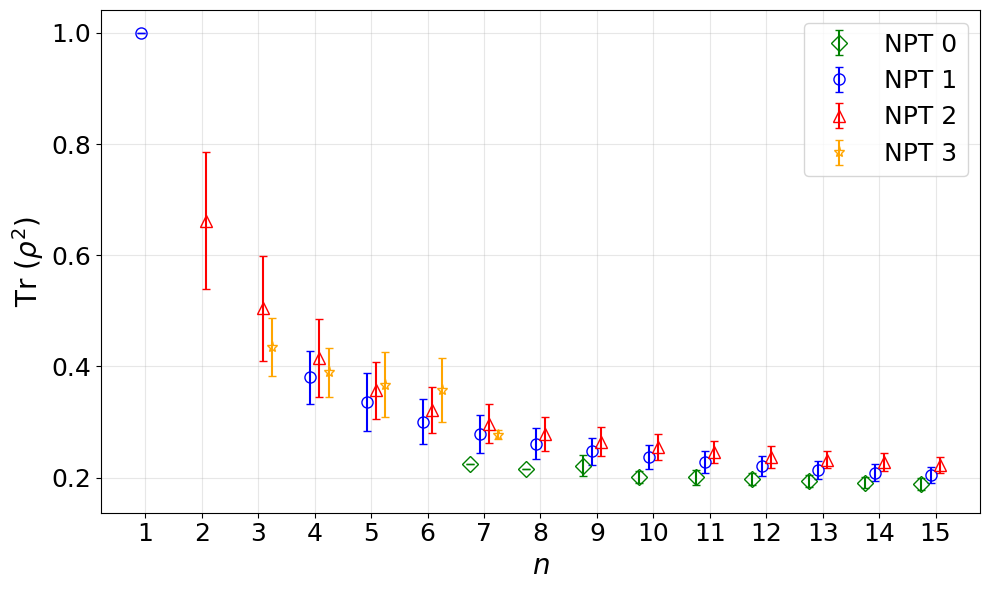}};
            \node[anchor=south west, xshift=5pt, yshift=5pt, fill=white, fill opacity=0.7, text opacity=1, inner sep=1pt] at (f1class.south west) {\subref*{fig:purity_mix_state}};
        \end{tikzpicture}
        \caption{}
        \label{fig:purity_mix_state}
    \end{subfigure}

    \hfill
    \captionsetup{justification=justified, singlelinecheck=false, format=plain, labelfont=bf}
    \caption{\justifying Summary of the mixed state method for generate quantum states. The number of sampled states per class depends on the value of $n$. For each $n$ 5000 samples have been generated.
    (a) Histogram for the mixed state method. 
    (b) Purity behavior for the different classes.}
    \label{fig:data_mixed_method}
\end{figure}

As all the techniques we employed for the characterization of states offering useful distillable quantum resources rely on machine learning, we require a large amount of data for the training process. It is infeasible to obtain states from a real quantum experiment, so we need to generate them. The data were generated synthetically: For each sample, a density matrix $\rho$ was constructed following one of the methods described below. The partial transpose $\rho^\Gamma$ was then computed with respect to the first subsystem, and the number of negative eigenvalues was evaluated in order to classify the corresponding quantum state. \\
We arbitrarily chose to set to zero the negative eigenvalues of the partially transposed density matrices whenever their magnitude was below $10^{-8}$, due to the lack of experimentally validated guidance on this threshold. This cutoff naturally affects the frequency with which states exhibiting a given number of negative eigenvalues are obtained; for example, increasing the threshold makes it more probable that a state with one negative eigenvalue  is generated respect to one with two negative eigenvalues.\\
We explored two different methods to generate random density matrices: 
In the first method, named mixed state method, the different states are taken as a convex combination of pure states:
    \begin{equation} \label{eq:mixed_state_method}
    \rho = \sum_\beta^n p_\beta \Ket{\psi_\beta} \Bra{\psi_\beta}, \quad \sum_\beta^n p_\beta = 1, \, p_\beta \geq 0 \, \, \forall \beta.    \end{equation}
Where the different $\ket{\psi_\beta}$ are entangled states sampled according to the Haar measure, namely columns of a random unitary matrix \cite{zyczkowski2001induced}, while the probabilities $\{p_\beta\}$ are sampled from the uniform distribution between zero and one and then normalized such that $\sum_{\beta} p_\beta = 1$.
The second method is named the Hilbert-Schmidt method. This is because it is the Hilbert-Schmidt measure that induces the joint probability distribution  in the simplex of the eigenvalues. It is briefly summarized by the following: 
\begin{equation}
    \hat{\rho}=\frac{\hat{A}\hat{A}^\dagger}{\text{Tr}(\hat{A}\hat{A}^\dagger)}
\end{equation}
where $\hat A$ is a random complex matrix whose real and imaginary entries are sampled from a normal distribution $\mathcal{N}(\mu=0, \sigma=1)$ \cite{zyczkowski2011generating}. \\
While both are good methods to generate a random density matrix, the number of states per class is not evenly distributed, as one can see in Fig. \ref{fig:hist_mix} and the second row of Table~\ref{tab:2x4_neg_eigs_montecarlo}. For the first method, it depends on the number $n$ of states in the mixture: we see that for a pure state ($n=1$) only NPT$_1$ states are generated, and then there is a complete transition for $n=2$ where all the states are NPT$_2$. When increasing $n,$ there is a little probability of generating NPT$_3$ states, but the majority are NPT$_2$ states. Then the probability shifts gradually from NPT$_2$ to NPT$_1$, and for higher values of $n$ there is also a non-zero probability of generating NPT$_0$ states. This is because states with high $n$ are very mixed, and the quantum character is smeared in favor of classical uncertainty. This can be observe by the purity, defined as the Tr$(\rho^2)$ behavior in Fig. \ref{fig:purity_mix_state}, where decrease when $n$ grows.   

\begin{figure}[h!]
    \centering

    \begin{subfigure}{0.5\textwidth}
        \captionsetup{labelformat=empty}
        \begin{tikzpicture}[remember picture]
            \node[inner sep=0] (f1ave) {\includegraphics[width=0.95\linewidth]{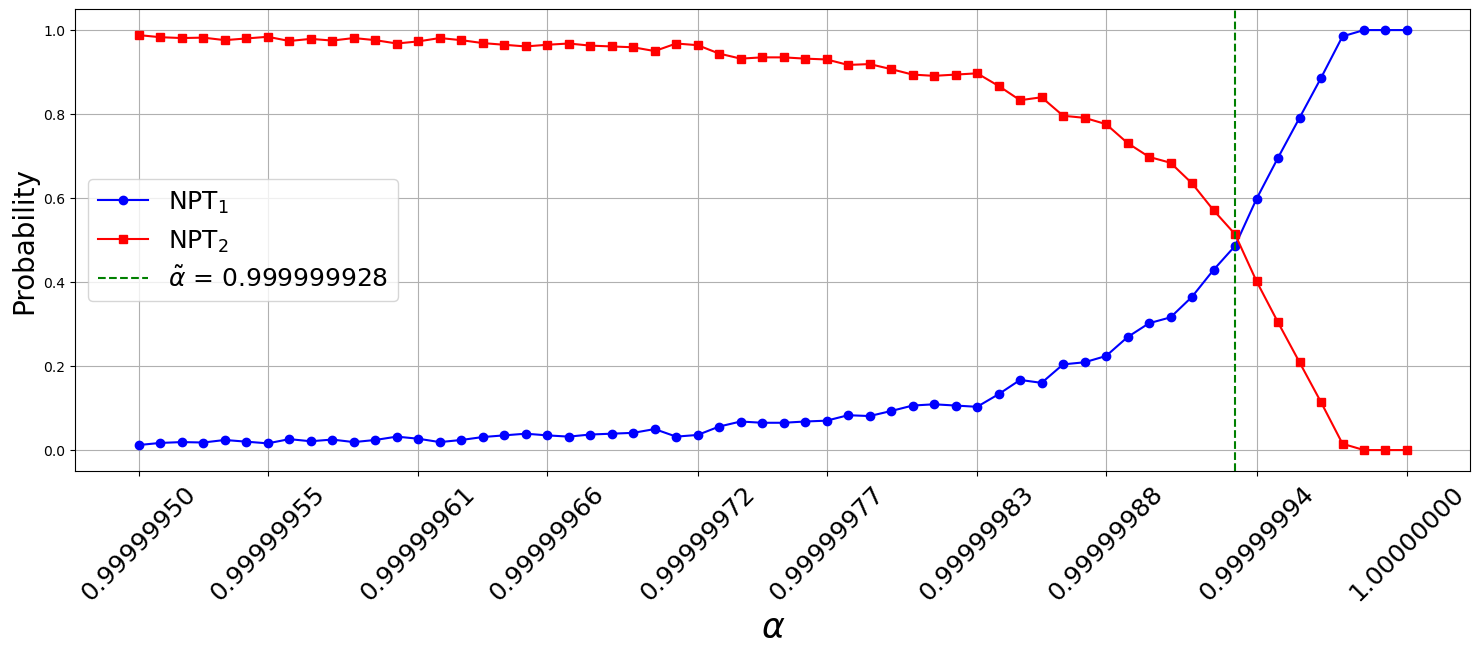}};
            \node[anchor=south west, xshift=5pt, yshift=5pt, fill=white, fill opacity=0.7, text opacity=1, inner sep=1pt] at (f1ave.south west) {\subref*{fig:transition_1_2}};
        \end{tikzpicture}
        \caption{} % Needed to increment subfigure counter
        \label{fig:transition_1_2}
    \end{subfigure}
    \hfill
    \begin{subfigure}{0.49\textwidth}
        \captionsetup{labelformat=empty}
        \begin{tikzpicture}[remember picture]
            \node[inner sep=0] (f1class) {\includegraphics[width=\linewidth]{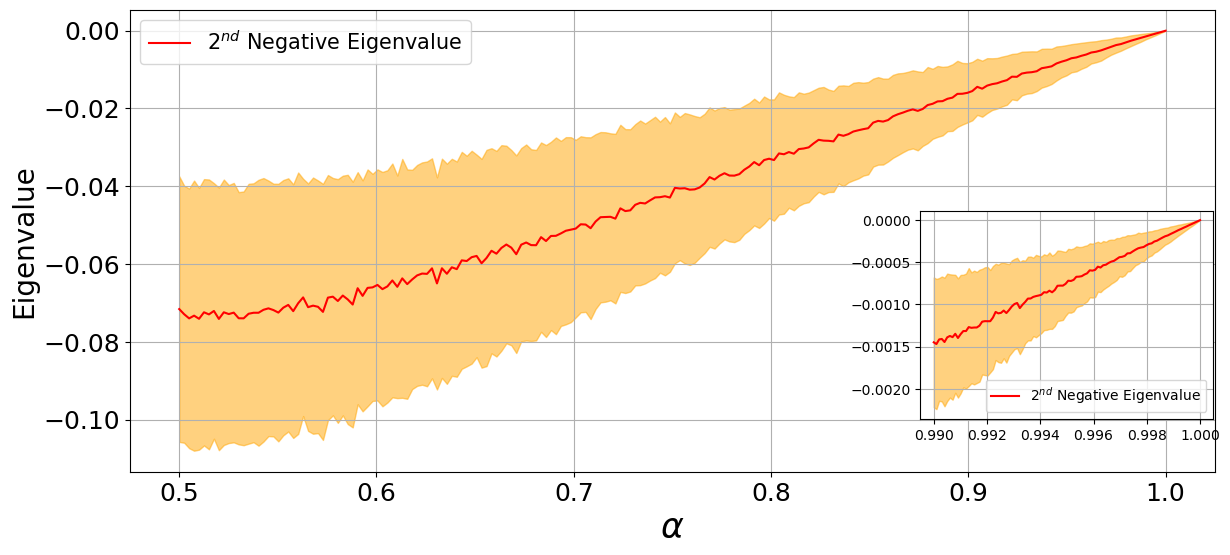}};
            \node[anchor=south west, xshift=5pt, yshift=5pt, fill=white, fill opacity=0.7, text opacity=1, inner sep=1pt] at (f1class.south west) {\subref*{fig:transition_value}};
        \end{tikzpicture}
        \caption{}
        \label{fig:transition_value}
    \end{subfigure}

    \hfill
    \captionsetup{justification=justified, singlelinecheck=false, format=plain, labelfont=bf}
    \caption{\justifying Analysis of the transition from NPT$_2$ to NPT$_1$ for $\alpha \to 1$ in Eq.~\ref{eq:mixture_of_two}.
    (a) Probability of sampling an NPT$_1$ or NPT$_2$ state as a function of $\alpha \to 1$ highlighting the point $\tilde{\alpha}$ where the probability is $50\%$ showing the transition from NPT$_2$ to NPT$_1$. 
    (b) Mean value with standard deviation of the $2^{nd}$ negative eigenvalue as a function of the parameter $\alpha$ in Eq.~\ref{eq:mixture_of_two}. To obtain the mean value, we sample 1000 density matrices for 200 values of $\alpha$ between $0.5$ and $1$ and we take the lower of the two negative eigenvalues after partial transposition. Inset is a zoom for $\alpha \in [0.990,1]$}
    \label{fig:transition}
\end{figure}

To better visualize the transition from $\xi=1$ to $\xi=2$ negative eigenvalues of $\rho^{\Gamma},$ we refer to Fig.~\ref{fig:transition_1_2}. 
Here we consider the state:
\begin{equation} \label{eq:mixture_of_two}
    \rho = \alpha  \Ket{\psi} \Bra{\psi} + (1-\alpha)\Ket{\phi}\Bra{\phi}
\end{equation} 
where $\ket{\psi}$ and $\ket{\phi}$ are entangled pure state sampled from Haar distribution and $\alpha \in [0,1]$.
We investigate the behavior of the state for values of $\alpha$ very close to $1$ in order to analyze the transition between NPT$_1$ and NPT$_2$ states. Around $\tilde{\alpha} = 0.999999928$, the probability of sampling an NPT$_1$ or an NPT$_2$ state is approximately $50\%$, as shown in Fig.~\ref{fig:transition_1_2}. This value depends on the numerical cutoff imposed on the second negative eigenvalue, which is set to $10^{-8}$ in our analysis. As highlighted in Fig.~\ref{fig:transition_value}, the second negative eigenvalue approaches zero as $\alpha \to 1$, indicating that the location of the transition point $\tilde{\alpha}$ is determined by the cutoff value: increasing the cutoff shifts the transition towards lower values of $\alpha$. A natural direction for future work would be to study experimentally the transition between NPT$_1$ and NPT$_2$ states, in order to determine physically reasonable values for the negative eigenvalues and to better justify the choice of the zero cutoff, which we currently set arbitrarily at $10^{-8}$.\\
As summarized in Table~\ref{tab:random_states}, very few states belonging to NPT$_0$ or NPT$_3$ are generated with standard random matrix generation. 
Given that, the metric-based methods can be more easily optimized to generate more states in less time, and overall, they do not impact the drawn conclusions.   We stress again that since the NPT$_3$ class can be omitted, as it does not significantly influence the classification, we choose this method for generating the datasets used in the training process.

\iffalse
    \begin{figure}[h!]
    \centering
    \includegraphics[width=0.4\linewidth]{images/Purity_diff_eig_HS.png}
    \caption{Plot of the distribution of the purity for states with a different number of negative eigenvalues.}
    \label{fig:pur2}
\end{figure}
\fi

\subsection{Collective measurements}
\label{app:cmw}
The first approach to the classification of $2\times4$ states is observables derived from the framework of Collective Measurement Witnesses (CMWs) \cite{rudnicki_collective_2011},  which can be seen as the probability of successfully projecting onto a singlet Bell state, given a particular pair of local projections. They can be represented by the following equation:
\begin{equation}
\label{eq:probabil}
    P_{xy} = \frac{\text{Tr}[(\hat{\rho}_T)(\hat{\Pi}_x\otimes\hat{\Pi}_{Bell}\otimes\hat{\Pi}_y)]}{\text{Tr}[(\hat{\rho}_T)(\hat{\Pi}_x\otimes\hat{I}\otimes\hat{\Pi}_y)]}
\end{equation}
where $\hat{\rho_T} = \hat{S}^T \hat{\rho} \hat{S} \otimes \rho$ with $\hat{S}$ the swap operator, $\hat{\Pi}_x$ and $\hat{\Pi}_y$ represent a projections onto a single qudit state, $\hat{\Pi}_{Bell}$ is a two-qubit projection onto singlet Bell state, and $\hat{I}$ denotes a four dimensional identity matrix.
The qudit projections are defined as $\hat{\Pi}_i = \ket{\psi_i} \bra{\psi_i},$ where $\{\ket{\psi_i}\}_{i\in\{1,..,16\}} $ are suitable SIC-POVM \cite{Bengtsson_2010}.
For a $2 \times 4$ system, there can be a total of 136 possible independent combinations of local projections.
These measurements are then used as input features for a neural network. The different configurations used of the measurement are report in table \ref{tab:02}.
\begin{table}[h!]
\centering
\caption{\justifying Definitions of the different sets of observables entering in Eq.~\eqref{eq:probabil}. From each configuration we construct $k$ observables to then use in our ML schemes. Note that we choose only the independent configuration so to exclude a simple exchange of local projectors.}
\label{tab:02}
\begin{tabular}{@{}cl@{}}
\hline
\textbf{$k$} & \textbf{Configurations} \\ \hline
1 & $\hat{\Pi}_1 \otimes \hat{\Pi}_1$ \\ \hline
8 & $\{\hat{\Pi}_i \otimes \hat{\Pi}_i\}_{i=1,\dots,8}$ \\ \hline
16 & $\{\hat{\Pi}_i \otimes \hat{\Pi}_i\}_{i=1,\dots,16}$ \\ \hline
32 & \begin{tabular}[t]{@{}l@{}}
        Configurations for B=16, plus the 16 additions: \\ 
        $\{\hat{\Pi}_i \otimes \hat{\Pi}_{i+1}\}_{i=1,\dots,15}$ and $\hat{\Pi}_{1} \otimes \hat{\Pi}_{16}$
     \end{tabular} \\ \hline
64 & \begin{tabular}[t]{@{}l@{}}
        Configurations for B=32, plus 32 varied additions, e.g., \\
        $\{\hat{\Pi}_i \otimes \hat{\Pi}_{i+2}\}_{i=1,\dots,14}$, $\{\hat{\Pi}_i \otimes \hat{\Pi}_{i+3}\}_{i=1,\dots,13}$, etc.
     \end{tabular} \\ \hline
136 & All 136 possible combinations \\ \hline
\end{tabular}
\end{table}

\begin{figure}[b!]
    %\centering
    \begin{subfigure}{0.49\textwidth}
        % --- ADD THIS LINE ---
        \captionsetup{labelformat=empty} 
        
        \begin{tikzpicture}[remember picture]
            % Place the image in a node
            \node[inner sep=0] (image1) {\includegraphics[width=\linewidth]{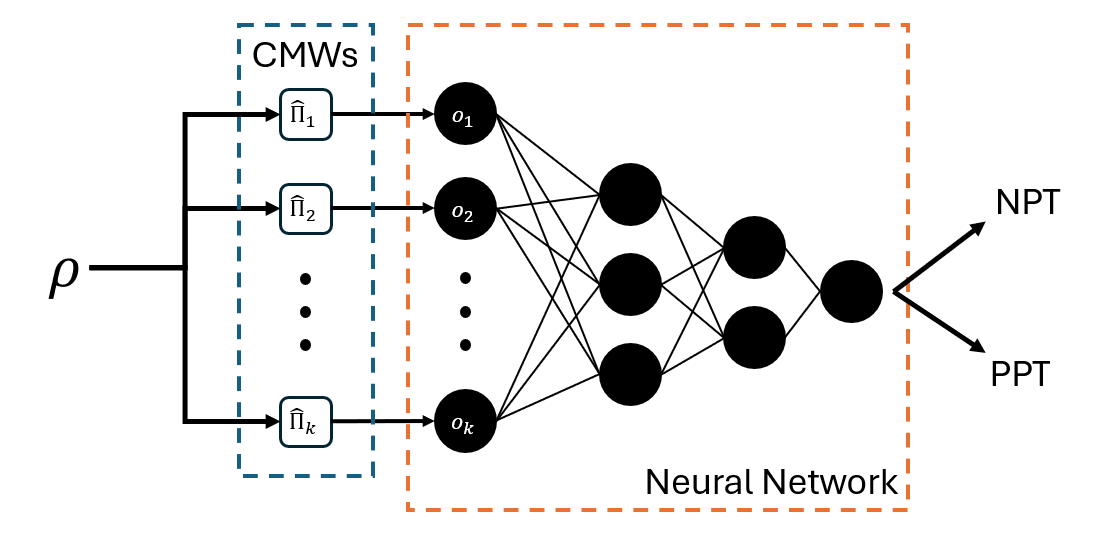}};
            % Manually place the label inside the image
            \node[anchor=south west, xshift=5pt, yshift=5pt, fill=white, fill opacity=0.7, text opacity=1, inner sep=1pt] at (image1.south west) {\subref*{fig:ill_1}};
        \end{tikzpicture}        
        \caption{} % Keep this empty caption to make the counter work
        \label{fig:ill_1}
    \end{subfigure}
    \hfill
    \begin{subfigure}{0.49\textwidth}
        % --- ADD THIS LINE ---
        \captionsetup{labelformat=empty} 
        
        \begin{tikzpicture}[remember picture]
            % Place the image in a node
            \node[inner sep=0] (image1) {\includegraphics[width=\linewidth]{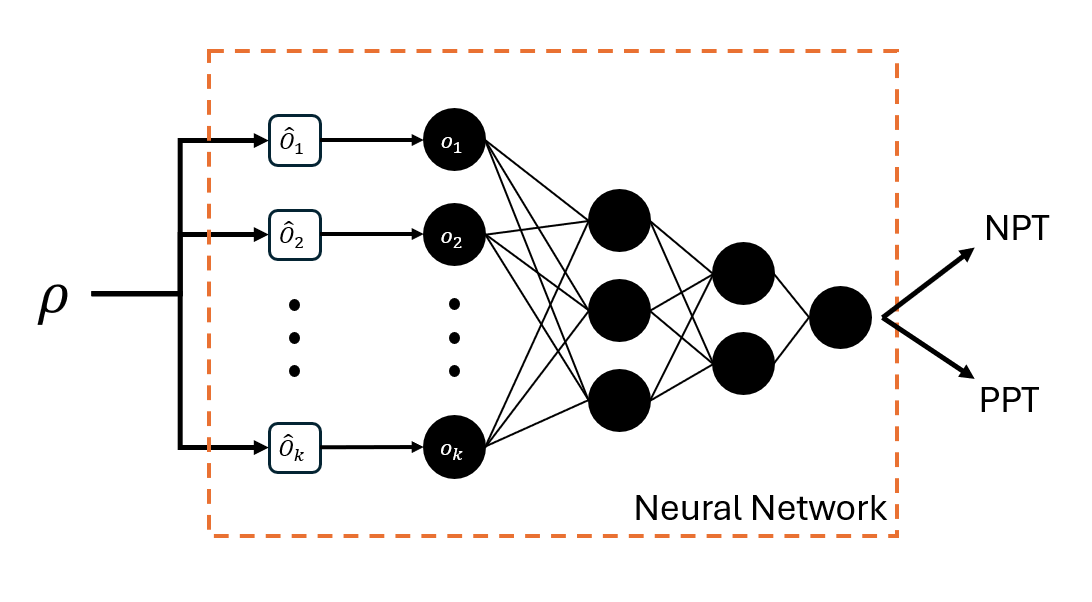}};
            % Manually place the label inside the image
            \node[anchor=south west, xshift=5pt, yshift=5pt, fill=white, fill opacity=0.7, text opacity=1, inner sep=1pt] at (image1.south west) {\subref*{fig:ill_2}};
        \end{tikzpicture}
        
        \caption{} % Keep this empty caption to make the counter work
        \label{fig:ill_2}
    \end{subfigure}
   \hfill
       \captionsetup{justification=justified, singlelinecheck=false, format=plain, labelfont=bf}
    \caption{\justifying
        Images illustrating the difference between the implemented methods for the classification of distillable states. 
       The primary distinction lies in how observables are integrated into the workflow: while the scheme in Figure (\subref{fig:ill_1}) utilizes a static set of observables, the scheme in Figure (\subref{fig:ill_2}) learns them dynamically during the training process. In both illustrations, the dashed orange rectangles denote the boundaries of the learning toolbox.
    }
    \label{fig:ill}
\end{figure}

\subsection{Learnable observables} \label{app:obs}
Another approach to classification is based on the idea that optimal observables may not be known a priori, but we can learn them through optimization. So, the idea is that our neural network model optimizes not only the weight but also the matrix entries of the observable used to calculate the features. A schematic representation of this method is illustrated in Fig. \ref{fig:ill}.

To do so, we add a custom layer to our neural network model, which is not composed of the usual weights but of the entries of $k$ complex matrices $\{\hat{A}_i\}$ that are randomly initialized. This first layer then does two things: firstly, it imposes user-defined conditions on the matrices $\{\hat{A}_i\}$. We imposed physics-informed constraints, namely that the observable needs to be Hermitian, and this can be simply implemented by the following formula:\begin{equation}\label{eq:obs1}
    \hat{O_i} = \frac{\hat{A_i}^\dagger + \hat{A_i}}{2} 
\end{equation}
subsequently the layer calculates the expectation values of the above observable by computing the trace:
\begin{equation}
    o_i = \operatorname{Tr}[\hat{O_i} \hat{\rho}].
\end{equation}
In the end, the set of features $\{o_i\}$ is passed to the neural network, which performs the final classification. We emphasize that all these calculations are differentiable, allowing the network's error to be backpropagated to all the way to the matrix elements of $\{\hat{A}_i\}$ using standard automatic differentiation.
From an experimental point of view, any such learned observable can, in principle, be implemented via a standard Naimark dilation \cite{busch1995operational}, i.e., by embedding the system into a larger Hilbert space with suitable auxiliary systems, applying a unitary, and measuring projectively on the extended space, as theoretically proposed in the context of single photon qudits \cite{Naimark_qudit} and applied in the context of quantum photonic processor \cite{Taballione_2019}.

The use of a single-copy measurement scheme is limited to functions that are linear in the density matrix $\rho$. However, entanglement indicators are generally nonlinear functions, such as the use of multiple copies of the density matrix enables the approximation of such functions \cite{PRL_2003}. This can be seen as a learnable generalization of the CMWs previously described.
This strategy is extended to the model described above, where the features and the input density matrix are computed on several replicas of the 
state under consideration, in the following way:
\begin{equation}
\label{eq:dm_cwm}
    o_i =  \operatorname{Tr} \left[ \hat{O}^{(l)}_i \hat{\rho}^{(l)} \right]
    \text{,} \quad \quad
    \hat{\rho}^{(l)} = \bigotimes_{i=1}^l \hat{\rho},
\end{equation}
where now $\hat{O}^{(l)}_i$ is a $(8l) \times (8l)$ observable obtain from the matrix $\hat{A_i}$ similarly to Eq.\eqref{eq:obs1}.

It is worth noting that with both methods, we have to choose some hyperparameters: the number of observables to calculate and the number of product states to use, in the case of the second technique. While the number of observable si taken similarly to those chosen for the CMWs method, as to more easily compare performance, the number of copies of the states is taken as a fixed $l=2$ because as $l$ grows, the computational intensity of the training process increases exponentially, rendering the calculation on our hardware infeasible, opening the way to exploitation of quantum machine learning strategies. 
In this work, the only condition imposed on the observables is that they must be Hermitian, as it represents the most general requirement for a physical measurement, but this is not the only choice. One direction would be to impose stricter conditions, such as taking the observable as projectors or enforcing some symmetry \cite{Torlai_2020}. Another strategy would be to investigate deeply the topological structure of the extracted data \cite{di_Pierro_2018}. Each option would define a different optimization landscape and, as such better solution may become available. Future investigations will systematically explore these alternative observable constraints, opening a compelling new line of research that bridges the geometry of quantum states with machine learning optimization.
%We leave the exploration of those alternatives for further study.

\subsection{NPT$_\xi$ classes under the lens of generalized Bloch decomposition}
\label{sec:svd}
We report here a few considerations about the negative eigenvalues and their use as resources for quantum information purposes. 
To such an extent, let us recall that we consider a bipartite state $\rho$ acting on $\mathcal{H}_A\otimes\mathcal{H}_B\cong\mathbb{C}^2\otimes\mathbb{C}^4$.
We denote by $\rho^\Gamma$ the partial transpose with respect to one subsystem (the spectrum is independent of which side is transposed).
A convenient canonical representation for a $2\times 4$ (qubit--ququart) state is the generalized Bloch/Fano form using tensor products of traceless Hermitian generators of $\mathrm{SU}(2)$ and $\mathrm{SU}(4)$ (e.g., Pauli matrices on the qubit and generalized Gell--Mann matrices on the ququart).  

\begin{figure*}
    \centering
    \includegraphics[width=1\textwidth]{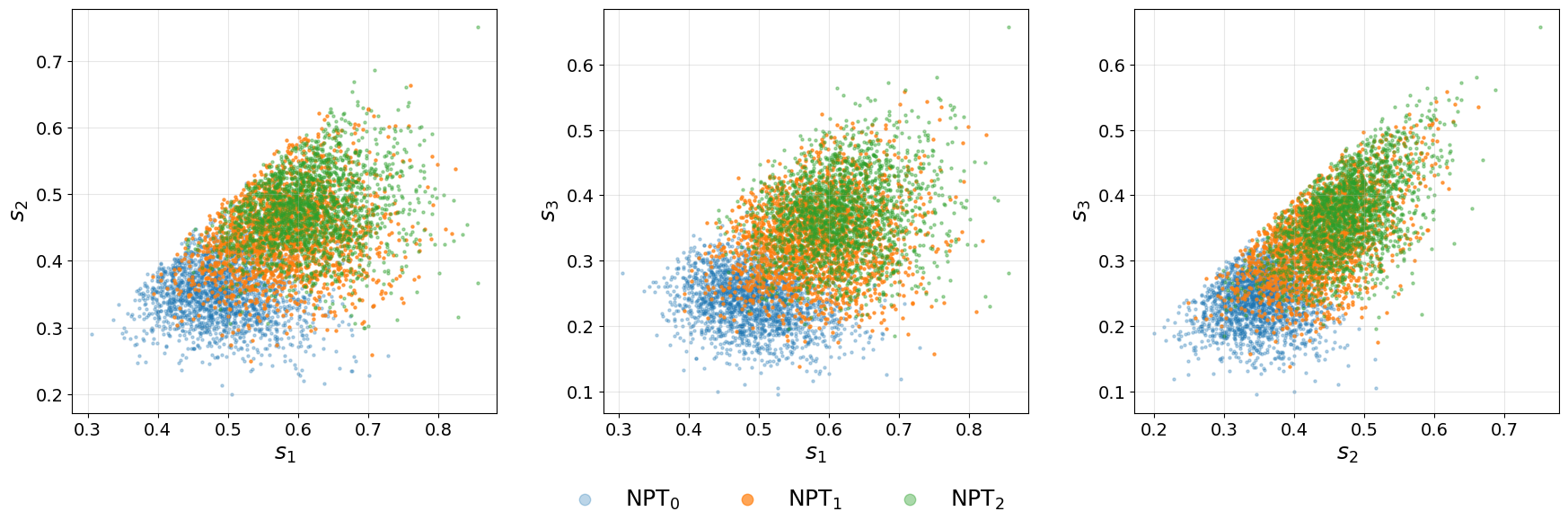}
    \hfill
    \captionsetup{justification=justified, singlelinecheck=false, format=plain, labelfont=bf}
    \caption{\justifying Three 2D projections of the singular values of the correlation matrix $T$ for quantum states from three different classes: NPT$_0$ (blue), NPT$_1$ (orange), and NPT$_2$ (green). The states are generated using the Hilbert-Schmidt method with 2000 samples per class. The panels show the projections $s_1$ vs $s_2$ (left), $s_1$ vs $s_3$ (center), and $s_2$ vs $s_3$ (right), where $s_1$, $s_2$, and $s_3$ are the singular values in descending order.}
    \label{fig:svd_bloch_decomposition}
\end{figure*}

Choose Hermitian operator bases $\{\sigma_i\}_{i=1}^3$ for $\mathfrak{su}(2)$ and $\{\Lambda_a\}_{a=1}^{15}$ for $\mathfrak{su}(4)$, normalized as 
$ \mathrm{Tr}(\sigma_i\sigma_j)=2\delta_{ij},\qquad \mathrm{Tr}(\Lambda_a\Lambda_b)=2\delta_{ab}, $
with $\mathrm{Tr}(\sigma_i)=\mathrm{Tr}(\Lambda_a)=0$, where the $\Lambda_a$ can be taken as generalized Gell--Mann matrices, 
which are traceless Hermitian and orthogonal in the Hilbert-Schmidt inner product \cite{KHANEJA200111}. 
%`\textcolor{red}{add reference article of the Gell-Mann matrices used ("Cartan decomposition of $SU(2^n)$ and control of spin system", Navin Khaneja)}
Then any density operator $\rho$ on $\mathbb{C}^2\otimes\mathbb{C}^4$ can be expanded uniquely as:
\begin{equation}
\rho
=\frac{1}{8}\Bigg( \mathbb{I}_2\otimes \mathbb{I}_4
+\sum_{i=1}^3 a_i\,\sigma_i\otimes \mathbb{I}_4
+\sum_{a=1}^{15} b_a\,\mathbb{I}_2\otimes \Lambda_a +\sum_{i=1}^3\sum_{a=1}^{15} T_{ia}\,\sigma_i\otimes \Lambda_a \Bigg),
\end{equation}
where $a\in\mathbb{R}^3$ and $b\in\mathbb{R}^{15}$ are local generalized Bloch vectors for the qubit A and the ququart B, while $T\in\mathbb{R}^{3\times 15}$ is the correlation matrix.
The coefficients are recovered by Hilbert-Schmidt projections: $ a_i=\mathrm{Tr}\!\left[\rho\,(\sigma_i\otimes \mathbb{I}_4)\right],\quad
b_a=\mathrm{Tr}\!\left[\rho\,(\mathbb{I}_2\otimes \Lambda_a)\right],\quad
T_{ia}=\mathrm{Tr}\!\left[\rho\,(\sigma_i\otimes \Lambda_a)\right].$\\
We perform some analysis using this decomposition to gain information about the geometry of the space of quantum states based on their number of negative eigenvalues under the PPT criteria. First, we consider the Singular Value Decomposition (SVD) of the correlation matrix $T$. Since $T$ is $3\times15$ in our case, the SVD gives in output 3 singular values (the minimum between 3 and 15). This singular $s_1, s_2$ and $s_3$ obey the relation $s_1 \geq s_2 \geq s_2$. They quantify the correlation between the two subsystems. We show them in Fig.~\ref{fig:svd_bloch_decomposition} where we project the 3D space in the three 2D projections. Here we consider 2000 states for each represented class (NPT$_0$, NPT$_1$, NPT$_2$) sampled with the Hilbert-Schmidt method. We observe a partial separation between the clusters corresponding to the three classes based on the magnitude of the singular value. In general, larger singular values are associated with more strongly entangled states, which are more likely to belong to the NPT$_2$ class rather than to the NPT$_0$ class. Conversely, smaller singular values correspond to less entangled states.\\
Nevertheless, these classes exhibit extensive overlap, meaning that such an analysis based on the singular-values of the correlation also offers provable evidence in the feature selection to distinguish among the three classes of states. The results corroborate our insight into the intricate structure of the distillable states in $d>6.$ 

\begin{figure*}
    \centering
    \includegraphics[width=1\textwidth]{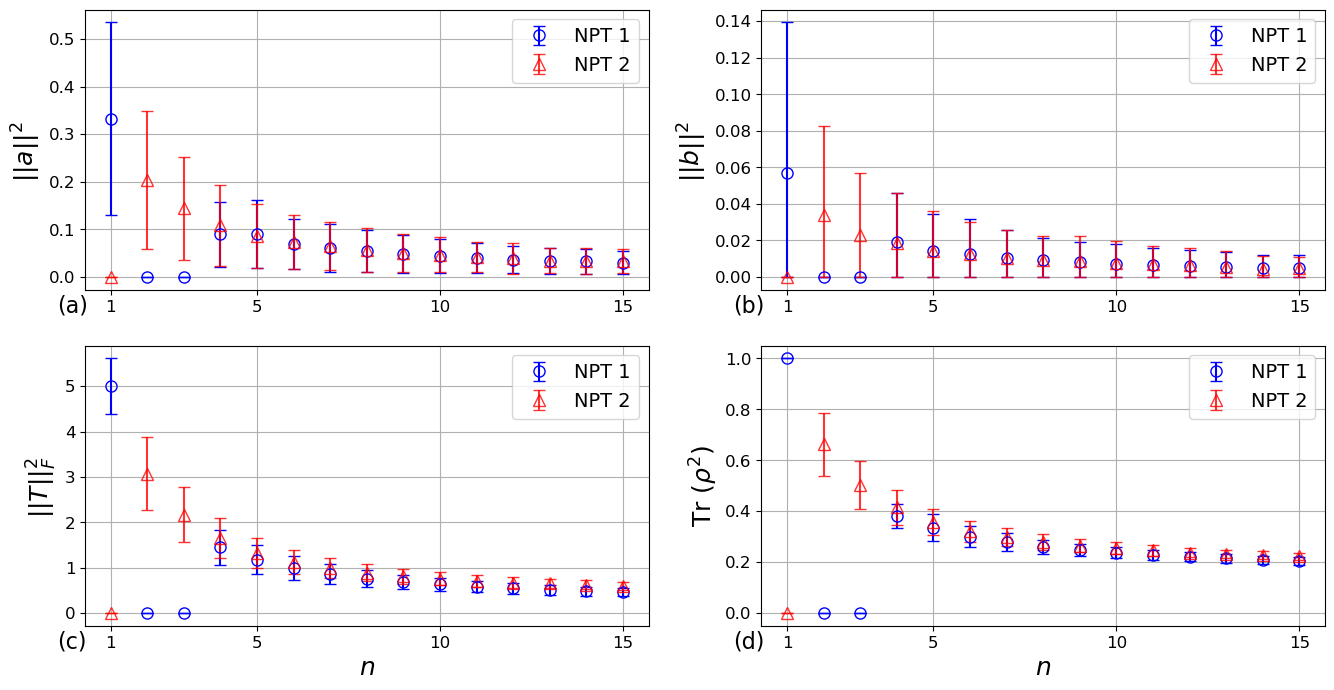}
    \hfill
    \captionsetup{justification=justified, singlelinecheck=false, format=plain, labelfont=bf}
    \caption{\justifying Bloch decomposition of mixed states with $n\in[1,15]$ in NPT$_1$ and NPT$_2$ classes. For each value of $n$ are generated 5000 samples and then are divided in the two classes. The panels (a) and (b) show the mean value of the Bloch vectors $\boldsymbol{a}$ and $\boldsymbol{b}$ for the two classes. Panel (c) shows the mean value of the Frobenius squared norm and panel (d) shows the mean value of the purity for the two different classes. For NPT$_1$ there are not states with $n=2,3$ so the mean values are set to 0; Same for NPT$_2$ for $n=1$}
    \label{fig:bloch_decomposition_analysis}
\end{figure*}

Another similar analysis was to consider the square norm of the Bloch vectors $\{a_i\}_{i=1}^{3}$ and $\{b_a\}_{a=1}^{15}$ and the Frobenius norm of the correlation matrix $T$ defined as  
\begin{equation}
    \|T\|_F = \sqrt{\sum_{i=1}^{m_1} \sum_{a=1}^{m_2} |T_{ia}|^2} = \sqrt{\text{Tr}(T^T T)},
\end{equation}
where in our case $m_1=3$ and $m_2=15$. We generate 5000 samples for values of $n\in [1,15]$ and according to the number of negative eigenvalues we divide the states in the classes. Then we compute the squared norms for the quantities mentioned before for each state in the different classes. The results are shown in Fig. \ref{fig:bloch_decomposition_analysis}. 
We restrict our analysis to states belonging to the NPT$_1$ and NPT$_2$ classes, as these represent the most relevant cases for the present study. The goal of this analysis is to assess whether the obtained quantities allow for a reliable discrimination between the two classes.
Figure~\ref{fig:bloch_decomposition_analysis}(a), (b), and (c) show the mean values of the squared norms of the Bloch vectors $\boldsymbol{a}$ and $\boldsymbol{b}$, and of the correlation matrix $T$, respectively. We observe that the Bloch vector associated with the qubit subsystem, $\boldsymbol{a}$, is generally one order of magnitude larger than the Bloch vector of the ququart subsystem, $\boldsymbol{b}$.
Figure~\ref{fig:bloch_decomposition_analysis}(d) reports the purity of the states, defined as $\mathrm{Tr}(\rho^2)$, for the two classes. At first inspection, a complete overlap between the two classes is observed for all the analyzed quantities, confirming the intrinsic difficulty in distinguishing between them using these descriptors alone.
Another notable observation is that the Bloch vectors and the correlation matrix exhibit the same qualitative behavior as the purity. This indicates that, for increasing values of $n$, the quantities arising from the Bloch decomposition reflect the increasing mixedness of the states, i.e., a higher degree of classical uncertainty.\\

\subsection{Consideration on the negative eigenvalues of the partial transpose in $2\times 4$}
\label{sec:2x4_pt_neg}
A basic limitation on the spectrum of $\rho^\Gamma$ is that the number of its strictly negative eigenvalues $\xi$ (counted with multiplicity) is upper bounded by:
\begin{equation}
\#\{\xi(\rho^\Gamma)<0\}\le (m-1)(n-1)
\end{equation}
for any $m\times n$ bipartite state $\rho$, therefore in our case they are limited to $\#\{\xi(\rho^\Gamma)<0\}\le 3.$
The proof mechanism behind the bound is geometric: if the negative eigenspace of $\rho^\Gamma$ has a dimension strictly larger than $(m-1)(n-1)$, then its span would necessarily contain a product vector.
A product vector in a negative subspace would imply a negative expectation value of $\rho$ on another (conjugated) product vector, contradicting $\rho\ge 0$.
As a consequence, whenever the bound is saturated, the negative eigenspace is a maximally complete entangled subspace (i.e., it contains no product vectors) \cite{rana_negative_2013}, while general separability criteria based on the correlation tensor have been pointed out in \cite{Scala_sep}.
The bound is tight in the qubit--qudit setting, as reported in \cite{Qubit_qudit_ppt}, one can construct for each $2\otimes N$ and each $m\in\{1,\dots,N-1\}$, explicit NPT states whose partial transpose has exactly $m$ negative eigenvalues. 
A numerical analysis of the qubit-ququart scenario is reported in table \ref{tab:random_states}, where we highlight how the \textit{typical} partial-transpose spectrum in $2\times4$  depends strongly on the chosen random-state ensemble and enforce the assumption discussed in \ref{sec:results}. In table \ref{tab:random_states}, we consider $N=10^6$ random realization of quantum states: i) pure radom states distributed according the Haar measure of $SU(8)$, ii) mixed random states with the measure induced by the Hilbert-Schmidt metric over the eigenvalues simplex, iii) as in the previous point but with the Bures metric replacing the HS one. 

As expected, pure Haar random states always exhibit exactly one negative eigenvalue of $\rho^\Gamma$. The ensemble in ii) is dominated by the case $\xi=1$ ($P(\xi=1)\approx 0.648$), with a substantial fraction at $\xi=2$ ($P(\xi=2)\approx 0.350$). PPT states ($\xi=0$) are already very non-typical ($P(\xi=0)\sim 10^{-3}$), and the maximally negative case $\xi=3$ is essentially negligible at the $10^{-6}$ level in our sample, indicating that spectra with three negative eigenvalues are atypical under HS sampling. In contrast, the Bures ensemble is clearly dominated by $\xi=2$ ($P(\xi=2)\approx 0.755$), while $\xi=1$ occurs much less frequently ($P(\xi=1)\approx 0.245$). The PPT case $\xi=0$ is extremely suppressed ($P(\xi=0)\sim 10^{-5}$), and $\xi=3$ appears at a small but visible rate ($P(\xi=3)\sim 4\times 10^{-4}$), making Bures sampling the ensemble with the highest observed incidence of the rare $\xi=3$ pattern. 

Now we briefly discuss how the number $\xi$ of strictly negative eigenvalues of $\rho^\Gamma$ (counted with multiplicity) impacts on a quantum distillation protocol. 
We define $\{|\psi_{-,\alpha}\rangle\}_{\alpha=1}^k$ be an orthonormal set of eigenvectors spanning the negative-eigenspace.
The corresponding rank-one projectors are $P_{-,\alpha}=|\psi_{-,\alpha}\rangle\!\langle\psi_{-,\alpha}|$, and it is sometimes useful to group them into the total projector $P_-:=\sum_{\alpha=1}^k P_{-,\alpha}$ onto the negative subspace.
 $\xi\in\{0,1,2,3\}$.
The geometric intuition behind the bound is also instructive: if the negative-eigenspace of $\rho^\Gamma$ were too large, its span would necessarily contain a product vector.
A product vector $|e,f\rangle$ in a negative subspace would have $\langle e,f|\rho^\Gamma|e,f\rangle<0$, which implies $\langle e^*,f|\rho|e^*,f\rangle<0$ and contradicts positivity of $\rho$.
In particular, when the upper bound is saturated, the negative eigenspace is forced to be ``product-vector free'' and can be viewed as a maximal completely entangled subspace.
An upper bound becomes practically meaningful once tightness is understood.

For applications, it is often desirable to compress the NPT structure into an effective two-qubit picture via local projection on Bob.
Let $P_B$ be a rank-two projector on $\mathbb{C}^4$ and define the projected (and renormalized) state:
\begin{equation}
\tilde\rho = \frac{(\mathbb{I}_A\otimes P_B)\,\rho\,(\mathbb{I}_A\otimes P_B)}{\mathrm{Tr}[(\mathbb{I}_A\otimes P_B)\,\rho]}.
\end{equation}
If $P_B$ can be chosen such that:
\begin{equation}
(\mathbb{I}_A\otimes P_B)\,|\psi_{-,\alpha}\rangle = |\psi_{-,\alpha}\rangle\qquad\forall\,\alpha=1,\dots,k,
\end{equation}
then the entire negative-eigenspace of $\rho^\Gamma$ sits inside an effective two-qubit Hilbert space $\mathbb{C}^2\otimes P_B\mathbb{C}^4\cong\mathbb{C}^2\otimes\mathbb{C}^2$, and the projectors $P_{-,\alpha}$ may be regarded as two-qubit projectors after restriction to this subspace.

A simple support-based criterion clarifies when such a simultaneous embedding is possible.
Any vector $|\psi\rangle\in\mathbb{C}^2\otimes\mathbb{C}^4$ can be written as:
\begin{equation}
|\psi\rangle = |0\rangle\otimes|u\rangle + |1\rangle\otimes|v\rangle,
\end{equation}
with $|u\rangle,|v\rangle\in\mathbb{C}^4$, so its Bob-support is contained in $\mathrm{span}\{|u\rangle,|v\rangle\}$, which has dimension at most two.
For multiple negative eigenvectors, define the joint Bob:
\begin{equation}
\mathcal{U}_B := \mathrm{span}\big\{\mathrm{supp}_B(\psi_{-,1}),\dots,\mathrm{supp}_B(\psi_{-,k})\big\}\subseteq\mathbb{C}^4.
\end{equation}
Then a single rank-two projector $P_B$ capturing \emph{all} negative eigenvectors exists if and only if $\dim\mathcal{U}_B\le 2$.
This becomes increasingly restrictive as $\xi$ grows: for $\xi=1$ the unique negative eigenvector $|\psi_-\rangle$ always admits such an embedding by taking $P_B$ onto $\mathrm{span}\{|u\rangle,|v\rangle\}$ in the decomposition $|\psi_-\rangle=|0\rangle|u\rangle+|1\rangle|v\rangle$, while for $\xi=2$ and $\xi=3$ a simultaneous embedding into a single two-qubit subspace is nongeneric and requires the different negative eigenvectors to share the same two-dimensional support on Bob.
In particular, the extremal case $\xi=3$ saturates the general limit $\xi\le 3$, is known to occur for valid $2\times 4$ states \cite{Qubit_qudit_ppt}, and at the same time tends to force the negative subspace to spread across Bob's four-dimensional space unless the state is specially structured.
Taken together, these results give a compact spectral viewpoint on qubit--ququart entanglement.
Rana's theorem provides the sharp ceiling $\xi\le 3$ \cite{rana_negative_2013}, Chen--Djoković show that the values $\xi\in\{1,2,3\}$ are all achievable in $2\times 4$ \cite{Qubit_qudit_ppt}, and the geometry of the negative eigenvectors determines whether one can (or cannot) localize the NPT character inside a single effective two-qubit subspace via a rank-two projection on Bob.

\begin{table}[t]
\centering
\caption{\justifying Empirical distribution of the number $\xi$ of negative eigenvalues of the partial transpose $\rho^\Gamma$ for random $2\times 4$ states.
Each row corresponds to $N=10^6$ samples from the indicated ensemble. The constraint $\xi\le 3$ follows from the general bound $(m-1)(n-1)$.}
\label{tab:2x4_neg_eigs_montecarlo}
\begin{tabular}{lrrrrr}
\hline
Ensemble &  $P(\xi=0)$ & $P(\xi=1)$ & $P(\xi=2)$ & $P(\xi=3)$ \\
\hline
Pure (Haar) & 0.00 &  1.00  & 0.00 & 0.00 \\
Mixed (Hilbert--Schmidt) & 0.001251 &  0.648388 &  0.350358 &  $3 \times 10^{-6}$ \\
Mixed (Bures) & $9 \times 10^{-6}$ & 0.244887 & 0.754664 & 0.00044 \\
\hline
\end{tabular}
\label{tab:random_states}
\end{table}
%\appendix
%
%
\section*{Author Contributions}
C.C. and P.D.R contributed equally to the paper, performed all numerical results, and wrote the first draft of the manuscript. 
M.G. conceived the research and supervised the ML strategies.
P.H. provided insights into the distillation protocols and guidance.
A.M. conceived the research idea, confirmed parts of the numerical results, supervised the research, and wrote the final draft of the manuscript.  
All authors contributed to the discussions and to the preparation of the manuscript.
\section*{Data and code availability}
All results reported in this work are based on \emph{synthetic} data generated numerically, i.e., randomly sampled qubit--ququart density matrices together with derived labels (PPT/NPT and the number of negative eigenvalues of the partial transpose) and measurement-feature datasets computed from these states.
No experimental datasets were used.
The code used to generate the synthetic states, compute labels and features, train the machine-learning models, and reproduce all figures and numerical results will be shared by the corresponding author upon reasonable request.
\section*{Competing Interests}
The authors declare no competing interests.
\section*{Acknowledgments}
AM acknowledges the IRA Programme, project no. FENG.02.01-IP.05-0006/23, financed by
the FENG program 2021-2027, Priority FENG.02, Measure FENG.02.01., with the support of
the FNP.
AM and MG acknowledge Angelo Bassi for fruitful discussions and the University of Trieste for the hospitality. MG is supported by CERN through the CERN Quantum Technology Initiative.
CG and PDR acknowledge Alessio D'Anna and Simone Brusatin for useful discussion.
%
%\bibliography{bib_full}
%

\end{document}